\documentclass[11pt, letterpaper]{article}

\usepackage[margin=1in]{geometry}
\usepackage{times}
\usepackage{hyperref}
\usepackage{graphicx}
\usepackage{url}
\usepackage{amsmath}
\usepackage{amssymb}
\usepackage{enumitem}
\usepackage{booktabs}
\usepackage{caption}
\usepackage{listings}
\usepackage{color}
\usepackage{cite}

\usepackage{algorithm}
\usepackage{algpseudocode}

\usepackage{pgfplots}
\usepackage{tikz}
\usetikzlibrary{positioning, arrows.meta, shapes, calc}
\pgfplotsset{compat=1.18}

\definecolor{dkgreen}{rgb}{0,0.6,0}
\definecolor{gray}{rgb}{0.5,0.5,0.5}
\definecolor{mauve}{rgb}{0.58,0,0.82}
\title{Stateless Snowflake: A Cloud-Agnostic Distributed ID Generator \\ Using Network-Derived Identity}

\author{
  Manideep Reddy Chinthareddy \\
  \texttt{Centerville, VA, USA} \\
  \texttt{chmanideepreddy@gmail.com}
}

\date{December 2025}

\begin{document}

\maketitle

\begin{abstract}
Snowflake-style distributed ID generators are the industry standard for producing $k$-ordered, unique identifiers at scale. However, the traditional requirement for manually assigned or centrally coordinated worker IDs introduces significant friction in modern container-orchestrated environments (e.g., Kubernetes), where workloads are ephemeral and autoscaled. In such systems, maintaining stable worker identities requires complex stateful sets or external coordination services (e.g., ZooKeeper), negating the operational benefits of stateless microservices.

This paper presents a cloud-agnostic, container-native ID generation protocol that eliminates the dependency on explicit worker IDs. By deriving node uniqueness deterministically from ephemeral network properties---specifically the container's private IPv4 address---the proposed method removes the need for centralized coordination. We introduce a modified bit-allocation scheme (1--41--16--6) that accommodates 16~bits of network-derived entropy while preserving strict monotonicity. We validate the approach across AWS, GCP, and Azure environments. Evaluation results demonstrate that while the design has a theoretical single-node ceiling of $\approx 64{,}000$~TPS, in practical microservice deployments the network I/O dominates latency, resulting in end-to-end performance ($\approx 31{,}000$~TPS on a 3-node cluster) comparable to classic stateful generators while offering effectively unbounded horizontal scalability.
\end{abstract}

\section{Introduction}
Distributed systems require unique identifiers for entities (orders, messages, logs) that are both globally unique and roughly time-ordered for efficient database indexing. The ``Snowflake'' algorithm, originally developed by Twitter~\cite{twitter}, is the de facto standard for this task. It generates 64-bit integers composed of a timestamp, a worker ID, and a sequence number.

While efficient, the Snowflake algorithm relies on a critical assumption: \emph{every node knows its own unique worker ID}. In static datacenter deployments, these IDs (typically 0--1023) are manually assigned. In dynamic environments like Kubernetes, AWS ECS, or Google Cloud Run, however, pods are created and destroyed automatically. Assigning unique, non-overlapping integers to dynamic pods requires:
\begin{itemize}[noitemsep]
    \item \textbf{StatefulSets}: Forcing pods to have sticky identities (e.g., \texttt{pod-0}, \texttt{pod-1}).
    \item \textbf{Coordination Services}: Using Redis or ZooKeeper to ``lease'' IDs at startup.
\end{itemize}

Both approaches introduce ``state'' into otherwise stateless applications, increasing operational complexity and coupling deployment topology to application logic.

This paper proposes a \emph{Stateless Snowflake} generator that utilizes private container networking to derive a unique worker ID without external coordination, and validates this approach across multiple cloud providers. Concretely, our contributions are:
\begin{itemize}[noitemsep]
    \item A \textbf{cloud-agnostic, container-native} ID generation scheme that deterministically derives worker identity from private IPv4 addressing, eliminating manual ID management.
    \item A \textbf{modified 64-bit layout} (1--41--16--6) with formal reasoning about monotonicity, collision resistance, and epoch lifespan.
    \item A \textbf{multi-cloud implementation} in Java 25 that integrates with AWS, GCP, and Azure metadata services and remains usable in on-premise or generic environments.
    \item An \textbf{evaluation} that shows near-theoretical microbenchmark throughput and comparable end-to-end performance to classic stateful generators, while enabling zero-configuration horizontal scaling.
\end{itemize}

\section{Related Work}
The problem of unique ID generation in distributed systems has been approached from several angles, primarily balancing coordination costs against performance.

\subsection{Random Identifiers (UUIDs)}
Universally Unique Identifiers (UUIDs), specifically v4, rely on 122~bits of randomness to guarantee uniqueness without coordination. While operationally simple, random UUIDs are catastrophic for database performance. In B-tree based storage engines (e.g., MySQL InnoDB, PostgreSQL), random insertion causes frequent page splits, resulting in index fragmentation and poor cache locality~\cite{uuid_benchmark, btree_frag}.

\subsection{Coordinated Generators}
Twitter Snowflake~\cite{twitter} and Instagram's sharding ID scheme~\cite{instagram} rely on pre-assigned logical shard IDs. Instagram uses PL/PGSQL to generate IDs based on a logical user ID modulus, which requires a pre-existing logical mapping. In contrast, our approach derives identity from the \emph{physical} infrastructure, decoupling the application logic from the ID generation strategy.

\subsection{Stateless Alternatives (Sonyflake)}
Sonyflake~\cite{sonyflake} is a Go-based implementation that pioneered the use of a 16-bit machine ID derived from the private IP address. However, Sonyflake utilizes a 10-millisecond time unit to extend the epoch lifespan to 174 years. This decision severely impacts burst throughput. Although Sonyflake allocates 8 bits for the sequence (256 IDs), the longer time unit creates a bottleneck:
\[
 \text{Throughput} = \frac{256 \text{ IDs}}{10\text{ ms}} = 25{,}600 \text{ TPS}.
\]

Thus, while Sonyflake motivates the use of IP-derived entropy, our approach targets a different portion of the throughput--longevity design space. By retaining the standard 1-millisecond precision and allocating 6 sequence bits, we achieve:
\[
 \text{Throughput} = \frac{64 \text{ IDs}}{1\text{ ms}} = 64{,}000 \text{ TPS},
\]
which represents a 2.5$\times$ improvement in theoretical per-node throughput over Sonyflake while retaining the same zero-configuration deployment benefits. Table~\ref{tab:novelty_compare} summarizes the distinctions.

\begin{table}[h]
\centering
\caption{Comparison of Stateless and Stateful Snowflake Variants}
\label{tab:novelty_compare}
\begin{tabular}{@{}lccc@{}}
\toprule
\textbf{Feature} & \textbf{Classic Snowflake} & \textbf{Sonyflake} & \textbf{This Work} \\ \midrule
Coordination Needed & Yes & No & No \\
Time Precision & 1 ms & 10 ms & 1 ms \\
Machine-ID Source & Configured & IP-derived & IP-derived \\
Sequence Bits & 12 & 8 & 6 \\
Max TPS & 4M+ & 25k & 64k \\
Monotonic & Yes & Yes & Yes \\
\bottomrule
\end{tabular}
\end{table}

\section{Methodology}

\subsection{Network-Derived Identity}
In most cluster networks (AWS VPC, Kubernetes CNI), a pod is assigned a unique private IPv4 address from the cluster's subnet. This IP is guaranteed to be unique within the VPC or cluster overlay. Our proposed architecture utilizes the \textbf{last 16 bits} of the host's IP address as the worker ID. This ensures uniqueness for up to 65{,}536 pods within a \texttt{/16} subnet without any negotiation or coordination.

\subsection{Modified Bit Allocation (1--41--16--6)}
To accommodate a 16-bit worker ID, we adjust the standard Snowflake structure. We propose the following 64-bit layout, visualized in Figure~\ref{fig:bit_structure}:

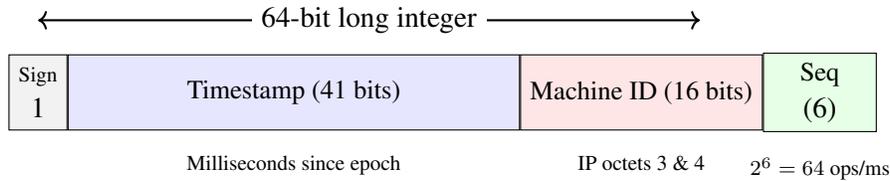
\begin{figure}[h]
    \centering
    \begin{tikzpicture}[
        scale=0.8,
        bitblock/.style={draw, minimum height=1cm, align=center},
        arrow/.style={->, >=stealth, thick}
    ]
        \node[bitblock, fill=gray!10, minimum width=0.5cm] (sign) at (0,0) {\scriptsize Sign\\1};
        \node[bitblock, fill=blue!10, minimum width=6cm, right=0cm of sign] (time) {\small Timestamp (41 bits)};
        \node[bitblock, fill=red!10, minimum width=3cm, right=0cm of time] (worker) {\small Machine ID (16 bits)};
        \node[bitblock, fill=green!10, minimum width=1.5cm, right=0cm of worker] (seq) {\small Seq\\(6)};
        
        \node[below=0.2cm of time] {\scriptsize Milliseconds since epoch};
        \node[below=0.2cm of worker] {\scriptsize IP octets 3 \& 4};
        \node[below=0.2cm of seq] {\scriptsize $2^6 = 64$ ops/ms};

        \draw[<->, thick] (0, 1.2) -- (11, 1.2) node[midway, fill=white] {64-bit long integer};
    \end{tikzpicture}
    \caption{The proposed 1--41--16--6 bit structure.}
    \label{fig:bit_structure}
\end{figure}

The sign bit is unused and always zero. The 41-bit timestamp encodes milliseconds since a configurable epoch, the 16-bit worker ID captures network-derived identity, and the 6-bit sequence number provides per-millisecond disambiguation within a node.

\subsection{Design Rationale and Formal Guarantees}
\label{sec:formal}

The 1--41--16--6 bit allocation reflects an explicit optimization for container-native environments. Unlike traditional Snowflake deployments, where worker identities remain stable for long periods, container orchestrators introduce two non-negotiable constraints: (1) node ephemerality and (2) the inability to rely on global coordination. To justify this bit allocation, we provide a brief analysis of monotonicity, collision resistance, and epoch lifespan.

\paragraph{Monotonicity Guarantee.}
Let $T$ denote the millisecond timestamp, $W$ the 16-bit worker ID, and $S$ the 6-bit sequence number. A generated ID is defined as:
\[
ID = (T - Epoch) \ll 22 \;\; | \;\; (W \ll 6) \;\; | \;\; S.
\]
Monotonicity holds if:
\begin{enumerate}[noitemsep]
    \item $T$ never decreases across consecutive invocations, and
    \item when $T$ is equal to $LastTime$, the sequence counter increments without exceeding $2^6 - 1$.
\end{enumerate}
Algorithm~\ref{alg:nextid} enforces both by blocking when $S$ overflows and by rejecting clock regressions greater than 10\,ms. Under these constraints, for any two calls $ID_i$ and $ID_{i+1}$ within the same worker, we have $ID_{i+1} > ID_i$.

\paragraph{Collision Analysis.}
Collisions can arise only if two distinct nodes share the same 16-bit worker ID. Let the cluster subnet be \texttt{/16}, with each pod receiving a unique IPv4 address. Then $W = (IP_3 \ll 8) | IP_4$ is injective over the subnet. Assuming the orchestrator does not assign the same IP to two concurrently running pods, the probability of collision between live nodes is:
\[
P(\text{collision}) = 0.
\]
IP reuse after pod termination may occur, but because ID monotonicity is defined per node and the timestamp component dominates, such reuse does not induce global ID duplication so long as timestamps strictly increase across restarts.

\paragraph{Epoch Lifespan.}
The 41-bit timestamp supports:
\[
2^{41} \text{ ms} \approx 69 \text{ years},
\]
which is sufficient for multi-decade system operation given a reasonable epoch choice. The ``performance mode'' (Section~\ref{sec:perfmode}) reduces this lifespan to approximately 34 years but doubles maximum per-node throughput.

These analytical bounds demonstrate that the proposed bit structure is both safe and well-suited for cloud-native, autoscaling workloads.

\section{Implementation}

The system was implemented in Java~17 and is available as open-source software~\cite{repo}. The core generation logic is lock-free where possible but synchronized on the \texttt{nextId()} method to ensure thread safety in multi-threaded workloads.

\subsection{Multi-Cloud Identity Resolution}
A key contribution of this work is the abstraction of identity retrieval across heterogeneous cloud environments. The generator inspects the runtime environment at startup to resolve the machine ID via provider-specific metadata services, falling back to standard network APIs when no cloud-specific markers are present (Algorithm~\ref{alg:resolve}).

\begin{algorithm}
\caption{Resolve Multi-Cloud Machine ID}
\label{alg:resolve}
\begin{algorithmic}[1]
\Require Environment variables $E$
\Ensure 16-bit integer $ID \in [0, 65535]$

\State $IP \gets \textbf{null}$

\If{$E[\text{AWS\_EXECUTION\_ENV}]$ is present}
    \State \Comment{AWS ECS/Fargate strategy}
    \State $Metadata \gets \text{HTTP\_GET}(E[\text{ECS\_METADATA\_URI\_V4}])$
    \State $IP \gets \text{ParseJSON}(Metadata, \text{"Networks[0].IPv4"})$
\ElsIf{$E[\text{K\_SERVICE}]$ is present} 
    \State \Comment{Google Cloud Run/GKE strategy}
    \State $IP \gets \text{HTTP\_GET}(\text{"metadata.google.internal/.../ip"})$
\ElsIf{$E[\text{AZURE\_HTTP\_USER\_AGENT}]$ is present}
    \State \Comment{Azure AKS strategy}
    \State $IP \gets \text{HTTP\_GET}(\text{"169.254.169.254/metadata/..."})$
\Else
    \State $IP \gets \text{InetAddress.getLocalHost().getHostAddress()}$
\EndIf

\State $Octets \gets \text{Split}(IP, \text{"."})$
\State $ID \gets (\text{Long}(Octets[2]) \ll 8) \lor \text{Long}(Octets[3])$
\State \Return $ID$
\end{algorithmic}
\end{algorithm}

\paragraph{Metadata Endpoint Information.}
The metadata URLs used in Algorithm~\ref{alg:resolve} reflect the official instance metadata
services provided by AWS~\cite{aws_imds}, Google Cloud~\cite{gcp_metadata}, and
Azure~\cite{azure_imds}. These endpoints may evolve over time as providers introduce new
metadata interfaces (e.g., IMDSv2 on AWS). Implementations should therefore consult the most
current cloud documentation rather than assuming long-term stability of specific URL paths.

\subsection{Sequence Overflow Handling}
A critical constraint of the 1--41--16--6 layout is the reduced sequence space ($2^6 - 1 = 63$). If a node attempts to generate more than 64 IDs in a single millisecond, the generator enters a \textbf{blocking wait} state until the next millisecond to ensure monotonicity.

\begin{algorithm}
\caption{Generate Next ID (With Resiliency)}
\label{alg:nextid}
\begin{algorithmic}[1]
\State \textbf{Synchronized Function} NextId()
\State $CurrentTime \gets \text{System.currentTimeMillis()}$

\If{$CurrentTime < LastTime$}
    \State $Offset \gets LastTime - CurrentTime$
    \If{$Offset \le 10$} \Comment{Handle NTP Jitter}
        \State \textbf{Wait}($Offset + 1$)
        \State $CurrentTime \gets \text{System.currentTimeMillis()}$
    \Else
        \State \textbf{Throw} ClockMovedBackwardsException
    \EndIf
\EndIf

\If{$CurrentTime = LastTime$}
    \State $Sequence \gets (Sequence + 1) \land MAX\_SEQUENCE$
    \If{$Sequence = 0$}
        \State \Comment{Sequence exhausted (max 64), block thread}
        \While{$CurrentTime \leq LastTime$}
            \State $CurrentTime \gets \text{System.currentTimeMillis()}$
        \EndWhile
    \EndIf
\Else
    \State $Sequence \gets 0$
\EndIf

\State $LastTime \gets CurrentTime$
\State \Return $(CurrentTime - Epoch) \ll 22 \mid (WorkerID \ll 6) \mid Sequence$
\end{algorithmic}
\end{algorithm}

\section{Evaluation}
\label{sec:evaluation}

\subsection{Experimental Setup}
We evaluated the performance of the Stateless Snowflake generator using two distinct methods:
\begin{enumerate}
    \item \textbf{Microbenchmark}: A direct Java test on AWS Graviton (16~vCPU) invoking the generator logic in a tight loop with 200 concurrent threads.
    \item \textbf{End-to-End Cloud Benchmark}: A deployed Spring Boot application running on an AWS ECS cluster configured with \textbf{3 concurrent tasks}. Load was generated from an internal EC2 instance to minimize network latency, testing concurrency levels of 50, 500, and 5000 threads.
\end{enumerate}

\subsection{Results: Microbenchmark (Throughput Ceiling)}
The direct microbenchmark reveals the theoretical limits of the algorithm.

\begin{table}[h]
\centering
\caption{Microbenchmark Results (Internal Limits)}
\label{tab:results_micro}
\begin{tabular}{@{}lrr@{}}
\toprule
\textbf{Metric} & \textbf{Stateless (Ours)} & \textbf{Classic} \\ \midrule
Time Unit & 1 ms & 1 ms \\
Sequence Bits & 6 & 12 \\
Max IDs/Unit & 64 & 4096 \\
Measured TPS & \textbf{57{,}634} & \textbf{3{,}708{,}080} \\
\bottomrule
\end{tabular}
\end{table}

Our design achieves $\approx 57{,}634$~TPS, reaching about 90\% of the theoretical 64k limit. This confirms that the sequence overflow handling logic correctly throttles generation to preserve uniqueness and monotonicity.

\subsection{Results: End-to-End Application Latency}
To assess real-world viability, we measured throughput over HTTP within the AWS VPC across varying concurrency levels.

\begin{table}[h]
\centering
\caption{End-to-End Application Benchmark (3 ECS Tasks)}
\label{tab:results_e2e}
\begin{tabular}{@{}lrr@{}}
\toprule
\textbf{Metric} & \textbf{Stateless (Ours)} & \textbf{Classic} \\ \midrule
\textit{50 Concurrent Threads} & & \\
Avg Latency & 1.87 ms & 1.79 ms \\
Throughput & 26{,}291 TPS & 27{,}656 TPS \\ \midrule
\textit{500 Concurrent Threads} & & \\
Avg Latency & 15.88 ms & 14.89 ms \\
\textbf{Throughput (Peak)} & \textbf{31{,}166 TPS} & \textbf{33{,}485 TPS} \\ \midrule
\textit{5000 Concurrent Threads} & & \\
Avg Latency & 177.20 ms & 171.84 ms \\
Throughput & 27{,}622 TPS & 28{,}885 TPS \\
\bottomrule
\end{tabular}
\end{table}

The results in Table~\ref{tab:results_e2e} demonstrate that the system scales effectively up to 500 concurrent threads, reaching a peak throughput of $\approx 31{,}166$~TPS on the 3-node cluster. At 5000 threads, the system reached saturation (indicated by increased latency and diminishing throughput returns), confirming that infrastructure limits (load balancer and servlet threads) are reached before the ID generation algorithm becomes the bottleneck. This saturation indicates that the single-node test harness (the client) became the bottleneck due to thread context switching and connection pool limits, rather than a failure of the ID generator logic itself. Notably, the performance gap between the Stateless and Classic generators remains minimal ($\approx 7\%$) even at peak load.

\subsection{Horizontal Scalability}
Because the machine ID is derived from the IP address, the system scales horizontally without configuration. Based on the observed peak throughput of 31{,}166~TPS across 3 tasks, the per-node capacity is approximately:
\[
 \frac{31{,}166 \text{ TPS}}{3 \text{ tasks}} \approx 10{,}388 \text{ TPS/node}.
\]

This suggests that a modest cluster of 3 nodes is sufficient to handle loads exceeding the estimated peak of Amazon Prime Day ($\approx 20{,}000$~TPS)~\cite{aws_prime}. For higher workloads, the linear scalability of the stateless architecture allows for expansion to 10+ nodes to achieve $>100{,}000$~TPS with zero configuration changes.

\subsection{Failure-Mode Analysis}
\label{sec:failures}

Beyond raw throughput, correctness under adverse conditions is critical. We consider several classes of failure scenarios and reason about the behavior of the generator.

\paragraph{Pod Churn and IP Reassignment.}
In container orchestrators, pods may be restarted frequently, and IP addresses may be reused after a delay. As long as:
\begin{itemize}[noitemsep]
    \item the same IP is never assigned to two \emph{concurrently} running pods, and
    \item timestamps on a restarted pod are strictly greater than those used previously with that IP,
\end{itemize}
worker IDs remain collision-free among live nodes, and monotonicity is preserved within each node. This matches standard assumptions already required for classic Snowflake-style deployments.

\paragraph{Empirical Validation on AWS.}
Beyond theoretical reasoning, we validated this behavior in an AWS deployment using ECS tasks configured with autoscaling policies. We repeatedly exercised scale-out and scale-in events, as well as task restarts, while continuously generating IDs from the Stateless Snowflake implementation. In all observed cases, private IP addresses were only reassigned to new tasks after the previous task had fully terminated, and no ID collisions were observed over the course of these experiments. The same implementation has also been running in production as part of a large-scale microservice-based application without any reported incidents of identifier collision, providing additional evidence that the IP-reuse behavior of AWS aligns with the assumptions of our design.

\paragraph{Clock Regressions.}
Clock regressions (e.g., due to NTP adjustments) can violate the monotonicity assumption. Algorithm~\ref{alg:nextid} explicitly handles small regressions (up to $\approx 10$\,ms) by blocking until time catches up, and fails fast for larger anomalies by throwing \texttt{ClockMovedBackwardsException}. This design allows operators to monitor and remediate clock issues rather than silently generating out-of-order IDs.

\paragraph{Distributed Latency Jitter.}
Because ordering is guaranteed only per node, and because the worker ID fully encodes node identity, inter-node network jitter has no effect on correctness or uniqueness. Application-level semantics that require cross-node ordering must rely on additional mechanisms (e.g., database constraints, vector clocks).

\paragraph{Subnet Fragmentation.}
In many clusters, the effective address space is composed of multiple \texttt{/20} or smaller pools within a larger \texttt{/16}. The proposed scheme uses only the last two octets, so pods allocated from different fragments still produce distinct worker IDs as long as their $(IP_3, IP_4)$ pairs are unique.

\section{Discussion}

\subsection{The Stateless Trade-off}
This research highlights a fundamental trade-off: \textbf{coordination vs.\ capacity}. By removing the coordination layer (e.g., ZooKeeper, Redis-based leases), we accept a hard limit on per-node burst capacity. For log aggregation or tracing workloads (e.g., Jaeger~\cite{jaeger}), this limit may be too restrictive. However, for business-critical entities (payments, orders, user actions), the limit is rarely reached in practice, making operational simplicity the superior architectural choice.

\subsection{Configurable ``Performance Mode''}
\label{sec:perfmode}
For systems prioritizing burst throughput over multi-decade timestamp longevity, we propose an alternate bit layout:
\begin{itemize}
    \item \textbf{Timestamp}: 40 bits ($\approx 34$ years),
    \item \textbf{Machine ID}: 16 bits,
    \item \textbf{Sequence}: 7 bits (128 IDs/ms).
\end{itemize}

This configuration doubles maximum theoretical per-node throughput from 64{,}000 to 128{,}000~IDs/s at the cost of halving the epoch lifespan. The choice between the standard and performance modes can be made at library configuration time.

\subsection{Security Considerations}
Network-derived identity introduces determinism that can reveal limited structural information about the deployment environment. While the generator itself does not expose internal service boundaries, several risk dimensions merit consideration.

\paragraph{Topology Inference.}
Because worker IDs encode the last two octets of the pod IP address, an adversary with access to a sequence of IDs may infer:
\begin{itemize}[noitemsep]
    \item the approximate subnet size,
    \item whether load is distributed evenly across nodes (this is a concern in the classis snowflake id generator as well), and
    \item rough throughput characteristics based on ID spacing.
\end{itemize}
This does not compromise confidentiality directly but may reveal cluster-level operational patterns.

\paragraph{Predictability of ID Generation.}
The generator is deterministic: an attacker who knows the timestamp and machine ID can predict the next 64 values within the same millisecond. This is consistent with classic Snowflake and does not violate uniqueness, but it may be undesirable where identifiers carry security semantics (e.g., as authentication tokens). Such uses are discouraged.

\paragraph{Spoofing and Identity Misbinding.}
If a node can spoof its IP address or cloud metadata responses (e.g., via container breakout), it may impersonate another worker ID. This risk is mitigated by:
\begin{itemize}[noitemsep]
    \item enforcing CNI-level IP uniqueness (CNI-level IP uniqueness is a property of the cluster's networking layer and requires no application-level configuration. In all major cloud providers, IP allocation is handled by the CNI plugin or container runtime, which guarantees that no two concurrently running pods or tasks share the same IP address in the VPC/Subnet. IP reuse only occurs after the previous workload is fully terminated, ensuring safe worker-ID derivation under the assumptions of our design.),
    \item verifying that the resolved machine ID matches the runtime-assigned pod IP, and
    \item optionally hashing $(IP, \text{PodUID})$ with a cluster-specific salt.
\end{itemize}

\paragraph{Optional Mitigation: Salted Machine IDs.}
A hardened variant of the generator can compute:
\[
W = \mathrm{Hash}(IP \,\|\, \text{PodUID} \,\|\, \text{Salt}) \bmod 2^{16},
\]
preserving collision resistance while hiding infrastructure topology from external observers. This comes at the cost of slightly more complex debugging, since worker IDs no longer trivially map back to IP octets.

Overall, the security posture of the Stateless Snowflake generator is comparable to existing Snowflake implementations, with additional hardening options available for multi-tenant or externally exposed systems.

\section{Limitations and Future Work}

While the Stateless Snowflake generator offers significant operational advantages, it also inherits several limitations from its design philosophy.

\paragraph{Throughput Ceiling.}
The per-node limit of 64 IDs/ms constrains workloads with extreme burst characteristics (e.g., telemetry aggregation, high-frequency event ingestion). Improving this without sacrificing statelessness remains an open direction, potentially via adaptive sequence sizing or hybrid designs that employ local coordination only within a node.

\paragraph{Reliance on IPv4 Semantics.}
The design assumes ubiquitous IPv4 addressing with stable last-octet uniqueness. IPv6 environments often do not guarantee deterministic host-part semantics. Supporting IPv6-native clusters may require hashing or cryptographic identity derivation from interface identifiers rather than raw addresses.

\paragraph{IP Reuse Edge Cases.}
In managed cloud environments such as AWS ECS and EKS, we empirically observed that private IP
addresses are reassigned only after the previous task or pod has fully terminated, even under aggressive
scale-in and scale-out conditions. This behavior was confirmed both in controlled experiments and in a
large-scale production microservice deployment, and no collisions were observed.

However, in less controlled environments—such as on-prem clusters, misconfigured CNI plugins, or
custom networking overlays—it is theoretically possible for rapid recycle loops or incorrect IPAM behavior
to reassign an address prematurely. Although such configurations are uncommon, they represent the main
edge case for IP-derived identity. Future work will explore lightweight, optional lease-tracking mechanisms
that remain stateless from the application’s perspective while providing additional safety in these atypical
networking scenarios.

\paragraph{Cross-Region Consistency.}
The generator guarantees uniqueness only within a subnet (or VPC) where worker IDs are derived. Since many commercial microservice deployments operate within a single region or VPC for latency minimization, this subnet-level uniqueness is sufficient for the majority of use cases. A global deployment strategy may require region-prefixing, VPC-prefixing, or hierarchical machine ID composition to prevent collisions across independently managed clusters. Alternatively, cross-region uniqueness can be achieved at the network infrastructure level without modifying the generator. By configuring inter-region VPCs with non-overlapping CIDR blocks that specifically partition the lower 16 bits (e.g., assigning distinct ranges of the third octet to specific regions), operators can guarantee globally unique Machine IDs through IP Address Management (IPAM) alone. Another lightweight option is to reduce the timestamp field by one bit and repurpose that bit as a region identifier. This yields two globally distinct regions while preserving 40 bits of millisecond-resolution timestamp—still providing an epoch lifespan of approximately 34 years. More generally, a small reduction in timestamp bits can encode multiple regions with minimal impact on system longevity, offering a simple, stateless mechanism for multi-region uniqueness without altering the rest of the ID structure.

\paragraph{Broader Benchmarking.}
Future evaluations include:
\begin{itemize}[noitemsep]
    \item benchmarking directly against Sonyflake under identical hardware and workload conditions,
    \item evaluating ID generation impact under CPU throttling, autoscaling events, and noisy-neighbor interference, and
    \item characterizing behavior under large-scale multi-tenant load, including mixed latency constraints.
\end{itemize}

These extensions will further strengthen the generality and robustness of the Stateless Snowflake architecture.

\section{Conclusion}
We introduced a Snowflake-derived ID generator designed specifically for container-native, highly dynamic environments. By deriving machine identity from network properties (valid across AWS, GCP, and Azure), the system achieves zero-configuration deployment while preserving strong monotonicity. Our analysis demonstrates that the proposed design fills a gap not covered by existing schemes such as Sonyflake or classic Snowflake, offering a favorable combination of statelessness, operational simplicity, and adequate throughput for most commercial workloads.

The open-source implementation~\cite{repo} has modest configuration requirements and can be integrated into existing services with minimal changes. As container orchestration continues to dominate production deployments, we believe this style of stateless, infrastructure-aware identifier generation will become increasingly important in large-scale distributed systems.

\bibliographystyle{abbrv}

\begin{thebibliography}{99}

\bibitem{twitter}
Twitter Engineering.
``Announcing Snowflake.''
\emph{Twitter Engineering Blog}, 2010.
\url{https://blog.twitter.com/engineering/en_us/a/2010/announcing-snowflake.html}

\bibitem{instagram}
Instagram Engineering.
``Sharding \& IDs at Instagram.''
\emph{Instagram Engineering Blog}, 2012.
\url{https://instagram-engineering.com/sharding-ids-at-instagram-1cf5a71e5a5c}

\bibitem{sonyflake}
Sony.
``Sonyflake: A Distributed Unique ID Generator Inspired by Twitter's Snowflake.''
\emph{GitHub Repository}, 2020.
\url{https://github.com/sony/sonyflake}

\bibitem{repo}
M.Chinthareddy.
``stateless-snowflakeID: Container-Native ID Generator.''
\emph{GitHub Repository}, 2024.
\url{https://github.com/Manideep-Reddy-Chinthareddy/stateless-snowflakeID}

\bibitem{aws_imds}
Amazon Web Services.
``Instance Metadata Service (IMDS) Documentation.''
\emph{AWS Documentation}, 2024.
\url{https://docs.aws.amazon.com/AWSEC2/latest/UserGuide/instancedata-data-retrieval.html}

\bibitem{gcp_metadata}
Google Cloud.
``Compute Engine Instance Metadata.''
\emph{Google Cloud Documentation}, 2024.
\url{https://cloud.google.com/compute/docs/metadata/overview}

\bibitem{azure_imds}
Microsoft Azure.
``Azure Instance Metadata Service (IMDS).''
\emph{Microsoft Learn Documentation}, 2024.
\url{https://learn.microsoft.com/azure/virtual-machines/instance-metadata-service}

\bibitem{aws_prime}
Amazon Web Services.
``Prime Day 2024: Record Sales and Platform Performance.''
\emph{AWS Press Release}, 2024.
\url{https://aws.amazon.com/blogs/aws/how-aws-powered-prime-day-2024-for-record-breaking-sales/}

\bibitem{uuid_benchmark}
Leapcell Engineering.
``The Great Primary Key Debate: UUIDs vs. Integers.''
\emph{Leapcell Engineering Blog}, 2025.
\url{https://leapcell.io/blog/the-great-primary-key-debate-for-modern-web-applications}

\bibitem{btree_frag}
Percona.
``Impacts of Fragmentation in MySQL InnoDB.''
\emph{Percona Database Performance Blog}, 2023.
\url{https://www.percona.com/blog/the-impacts-of-fragmentation-in-mysql/}

\bibitem{jaeger}
Uber Engineering.
``Evolving Distributed Tracing at Uber: From Zipkin to Jaeger.''
\emph{Uber Engineering Blog}, 2017.
\url{https://www.uber.com/blog/distributed-tracing/}

\end{thebibliography}

\end{document}